\documentclass[preprint]{revtex4-1} 
\usepackage{graphicx}
\usepackage[]{color}
\usepackage{amsmath,amssymb}
\usepackage{subfig} 
\newcommand\abs[1]{\left|#1\right|}
\usepackage{float}
\usepackage{xcolor}
\usepackage{chemformula}
\usepackage{amssymb}
\usepackage{mathtools}
\usepackage{float}
\newcommand{\ba}{\begin{eqnarray}}
\newcommand{\ea}{\end{eqnarray}}
\newcommand{\be}{\begin{equation}}
\newcommand{\ee}{\end{equation}}

\usepackage{soul}
\begin{document}
\author{Amin Bakhshandeh}
\affiliation{Instituto de F\'isica, Universidade Federal do Rio Grande do Sul, Caixa Postal 15051, CEP 91501-970, Porto Alegre, RS, Brazil.}

\author{Alexandre P. dos Santos}
\affiliation{Instituto de F\'isica, Universidade Federal do Rio Grande do Sul, Caixa Postal 15051, CEP 91501-970, Porto Alegre, RS, Brazil.}
\author{Yan Levin}
\email{levin@if.ufrgs.br}
\affiliation{Instituto de F\'isica, Universidade Federal do Rio Grande do Sul, Caixa Postal 15051, CEP 91501-970, Porto Alegre, RS, Brazil.}
\title{Interaction between charge-regulated metal nanoparticles in an electrolyte solution}
	
	\begin{abstract}
		
		We present a theory which allows us to calculate the interaction potential between charge-regulated metal nanoparticles inside an acid-electrolyte solution.  The approach is based on the recently introduced model of charge regulation which permits us to explicitly --- within a specific microscopic model --- relate the bulk association constant of a weak acid to the surface association constant for the same weak acid adsorption sites.  When considering metal nanoparticles we explicitly account for the effect of the induced surface charge in the conducting core.  To explore the accuracy of the approximations, we compare the ionic density profiles of an isolated charge-regulated metal nanoparticle with explicit Monte Carlo simulations of the same model.  Once the accuracy of the theoretical approach is established, we proceed to calculate the interaction force between two charge-regulated metal nanoparticles by numerically solving the Poisson-Boltzmann equation with charge regulation boundary condition.   The force is then calculated by integrating the electroosmotic stress tensor.   We find that for metal nanoparticles the charge regulation boundary condition can be well approximated by the constant surface charge boundary condition, for which a very accurate Derjaguin-like approximation was recently introduced.  On the other hand, a constant surface potential boundary condition often used in colloidal literature, shows a significant deviation from the charge regulation boundary condition for particles with large charge asymmetry.
		
	\end{abstract}
	
	\maketitle
	
	\newpage
	
	\section{Introduction}
	Aqueous electrolyte and acid solutions are of great importance in chemistry and biology~\cite{lynden2010water,ZHONG20111,Trogu,Ball2013}. For strong acids and electrolytes, the high dielectric constant of water favors dissociation of ionic components~\cite{pliego2000,andelman2006}. The  ionization process plays a crucial role in many physicochemical phenomena, such as stability of colloidal and nonoparticle suspensions~\cite{Borkovec1999,Markovich_2016}.
	For example, the surface of silicate glass acquires a negative charge when immersed in water, which is related to  the dissociation of silanol groups~\cite{Behrens2001}. This and other examples of surface charging are described by the charge regulation~(CR) process which controls the exchange between dissociable acidic or basic functional groups with their conjugate electrolyte~\cite{Nap2014,linderstrom,podgornik2018,frydel2019,avni2019charge,majee2019,netz2002,Krishnan1,Krishnan2,monica1,monica2,Ong}. As the result, the behavior of these systems can change dramatically as a function of solution pH~\cite{Guzman}.  A particularly important example of charge regulation occurs in metal nanoparticles.    Gold nanoparticles are often synthesized using citrate as a
	stabilizing agent and their surface charge is a strong function of pH. The plasmon resonance of these particles has been exploited in sensors and optical devices~\cite{Eustis,Schaadt,hutter2}.  They have also found use in catalysis~\cite{Fumitaka,Alain,Taehoon,Ganganath,Pablo,Didier}. Because of their compatibility with the immune system, gold nanoparticles are now being used extensively in medical applications~\cite{Larese2009,Nassar2011,Chih-ping,Zhao}.  
	
\begin{figure*}[t]
		\begin{center}
			\includegraphics[scale=0.3]{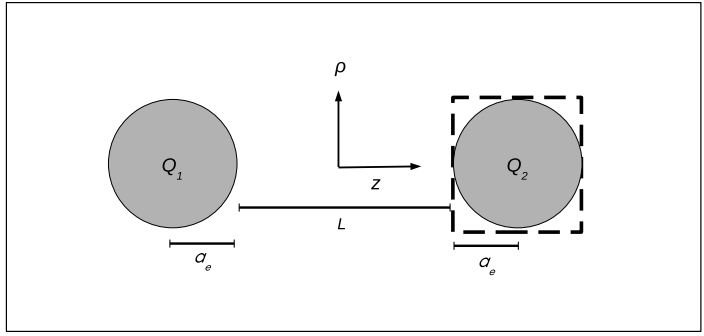}
		\end{center}
		\caption{Representation of the system. Two metal nanoparticles of charge $Q_1$ and $Q_2$ and effective radius $a_e=a+r_i$, where $a$ is the radius of nanoparticle and  $r_i$ is the ionic radius,  are separated by the contact surface-to-contact surface distance $L$, in an electrolyte-acid solution of respective concentrations $\rho_a$ and $\rho_s$. The origin of the coordinate systems is at the mid plane between the particles. The dashed box around the second particle corresponds to the cylindrical surface used to calculate the interaction force between the particles using the electroosmotic stress tensor.}
		\label{fig1}
	\end{figure*}

	The CR was first introduced by Linderstr{\o}m-Lang~\cite{linderstrom} and some years later Kirkwood and Shumaker explored the effects of CR on intermolecular interactions between proteins~\cite{kirkwood1952}. In recent years many studies have been conducted on proteins using CR~\cite{Lund2013,Blanco2019,Mikaellund,SHEN200592,Tsao2000,Jyh-Ping,Burak,Kumar,Krishnan3,markovich2018chapter,popa2010}. In $1971$ Ninham and Parsegian(NP)~\cite{ninham1971} quantitatively  implemented CR  for planar non-polar surfaces by combining the  local chemical equilibrium  of functional groups on the surface with  the  non-linear Poisson-Boltzmann equation~\cite{gouy1910,chapman19}. 
	The fundamental parameter in the NP approach is the association/dissociation equilibrium constant for acid or base surface functional groups, which is assumed to be the same as the bulk equilibrium constant for the same acid or base.  In the later works, the surface association constant was treated as an adjustable parameter used to fit the experimental data~\cite{behrens}.   
	NP approach is a mean-field theory which does not account for the discrete nature of ions and of the surface functional groups.  To understand the effects of discreteness one needs to consider an explicit microscopic model of surface association.  Such approach was recently introduced in Refs.~\cite{bakhshandeh2019,bk2020}.  The authors of these papers used the Baxter model of sticky spheres~\cite{Baxter} to account for the chemical association.  
	The fundamental conclusion of these works was to demonstrate that the discreteness effects lead to the 
	renormalization of the surface association constant away from its bulk value.  Within the specific model of association, the new theory provided an explicit mapping between the bulk and the surface equilibrium constants.   One can, therefore, accurately account for both the electrostatic and steric discreteness effects within the NP framework by a suitable renormalization of the bulk association constant.
	In the present paper we will explore the effects of charge regulation on the interaction between metal nanoparticles~\cite{DosSantos2019}, see Fig \ref{fig1}.
	
	\section{The discreteness effect of charged functional groups on metal surface}\label{A1}
	
\begin{figure*}[t]
		\begin{center}
			\includegraphics[scale=0.5]{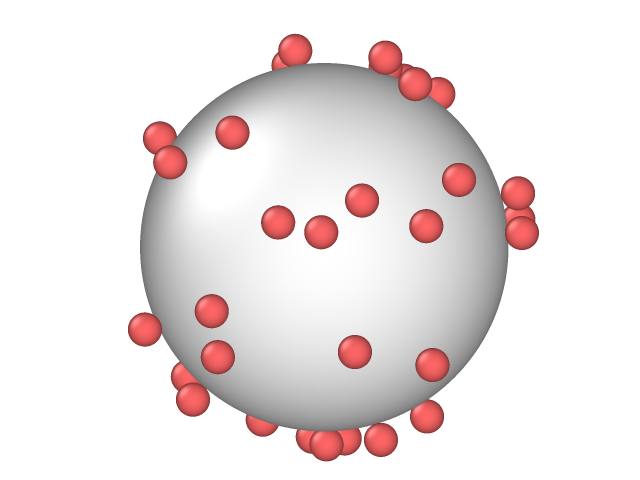}
		\end{center}
		\caption{Representation of a metal nanoparticle with Baxter sticky sphere, negatively charged adsorption sites, randomly distributed on its surface.}
		\label{fc}
	\end{figure*}

	Consider a metal nanoparticle of radius $a$ and $N_{s}$  acidic charged functional groups on its surface.  The nanoparticle is  placed at the 
	center of a spherical Wigner-Seitz (WS) cell of radius $R$, see  Fig.~\ref{fc}. 
	
	The system is in contact with a reservoir of strong acid at concentration $\rho_a=10^{-\rm pH}$M  and of strong electrolyte at concentration $\rho_s$.   The ionization equilibrium of surface groups is controlled by the  equilibrium constant $K_{s}$:
	\begin{equation}    
	\ch{H_3 O+}  + \ch{A-}    \rightleftarrows  \ch{HA} +\ch{H_2 O} \ .
	\label{Eq1}
	\end{equation} 
	The electrostatic potential $\phi$ inside the cell satisfies the Poisson equation in spherical coordinates: 	
	\begin{equation}  
	\frac{\epsilon_w}{r^2}\frac{\partial }{\partial r}\left( r^2 \frac{\partial \phi}{\partial r} \right) = -4 \pi\sigma_{eff} \delta(r-a_e)  - 4 \pi q \left[ \rho_+(r)- \rho_-(r) \right] ,
	\label{eq1}
	\end{equation}  
	where $a_e=a+r_i$ is the radius of the contact sphere --- the closest distance that  an ion of radius $r_i$ can reach the nanoparticle center.    $\epsilon_w$ is the dielectric constant of water, $q$ is the proton charge, and $\rho_\pm$ are the concentration of positive and negative ions. For monovalent ions electrostatic correlations between the ions can be neglected and the local concentration of cations and anions is given by the Boltzmann distribution~\cite{Levin2002}.  Eq. (\ref{eq1})  then reduces to the Poisson-Boltzmann equation,
	\begin{equation}
	\frac{\epsilon_w}{r^2}\frac{\partial }{\partial r}\left( r^2 \frac{\partial \phi}{\partial r} \right) = -4 \pi\sigma_{eff} \delta(r-a_e) + 8 \pi q (\rho_a+\rho_s) \sinh{[\beta q \phi]} \,,
	\label{eq1a}
	\end{equation}
	with the  effective surface charge given by~\cite{bakhshandeh2019,bk2020}
	\begin{equation}  
	\sigma_{eff} = - \frac{N_{s} q}{4 \pi a_e^2}\frac{1}{\left(1+ K_{s} \rho_a \mathrm{e}^{-\beta \phi_0}\right)}
	\label{Eq2}
	\end{equation}
	where $\beta=1/k_b T$ and  $\phi_0$ is the contact electrostatic potential. Within the simple model in which both the surface adsorption sites and the hydronium ions are described by the equal-sized sticky spheres of radius $r_i$, the surface equilibrium constant $ K_{s}$ was shown to be related to the bulk equilibrium constant of the same weak acid~\cite{bakhshandeh2019,bk2020}, $K_{bulk}$, as
	\begin{equation}\label{ksur}
	K_{s} = \frac{K_{bulk}}{2}\mathrm{e}^{-b-\beta \mu_d} \ ,
	\end{equation}
	where $b=\lambda_B/2 r_{i}$  and  $\lambda_B=\beta q^2/\epsilon_w$ is the Bjerrum length, which is $7.2$~\AA\ in water at room temperature.   The term $\mu_d$ accounts for the discrete nature of the surface acidic groups.  
	
	\begin{figure}[h]
		\begin{center}
			\includegraphics[scale=0.4]{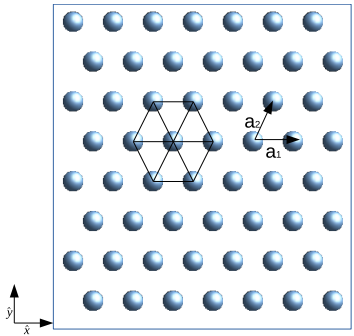}\vspace{0.2cm}\hspace{0.1cm}
		\end{center}
		\caption{Representation of a triangular lattice with their lattice vectors. }
		\label{triangular}
	\end{figure}
	The value of $\mu_d$ can be obtained by considering the interaction of an associated hydronium ion with all the surface acid groups, as well as,  with its own self-image, and with the images of the acid groups inside the metal core. To calculate $\mu_d$, for simplicity, we will neglect the curvature of the particle surface.  This is a reasonable approximation as long as the radius of the nanoparticle is much larger than the average separation between the surface charged groups~\cite{Rouzina,Arenzon2002,lau2001,levin2003}. 
	Furthermore, we will assume that the charged sites form a triangular lattice over the particle surface, as is shown in Fig.~\ref{triangular}.  In reality, the acidic functional groups are randomly distributed over the surface, however, we find that the precise site arrangement does not play a significant role in CR, so that triangular lattice distribution provides a reasonable first order approximation.  
	Since a nanoparticle is metallic, each site crates an image charge at the inversion point inside the metal core.   
	We note that the mean field potential  $\phi$ already accounts for the uniform -- smeared -- surface charge, therefore, it is necessary to  introduce a uniform neutralizing background to prevent the double counting~\cite{bakhshandeh2019,bk2020}.   The $\mu_d$ can then be separated into the contribution arising from the direct interaction between the hydronium ion and the negatively charged adsorption sites and their neutralizing background, $\mu_s$; and  the contribution arising from the interaction of the hydronium with its own image inside the metal core, with the images of the adsorption sites and the uniform neutralizing image background, $\mu_i$. 
	\begin{equation}
	\mu_d = \mu_s +\mu_i \ ,
	\label{mut}
	\end{equation}
	
	The average contact surface area per functional group is  $\gamma = 4 \pi a_e^2 /N_{s}$.  Since the unit cell of a triangular lattice has area $\gamma = \frac{\sqrt{3}}{2} l^2$, where $l$ is the lattice spacing, the separation between the surface groups is 
	found to be:
	\begin{equation}
	l=\frac{2 a_e}{3^{\frac{1}{4}}}\sqrt{\frac{2 \pi}{N_{s}}}.
	\end{equation}
	The lattice vectors for the planar, $z=0$, triangular lattice of adsorption sites are:
	\begin{equation}
	\begin{split}
	\boldsymbol{\mathrm{a}}_1 = l~{\boldsymbol{\hat{\mathrm x}}} \ \ \ \ \text{and} \ \ \ \
	\boldsymbol{\mathrm{a}}_2 = \frac{1}{2} l~ {\boldsymbol{\hat{\mathrm x}}} + \frac{\sqrt{3}}{2} l~{\boldsymbol{\hat{\mathrm y}}} \ .
	\end{split}
	\label{vector}
	\end{equation}
	Consider a hydronium ion of charge $q$ in contact with the site of charge $-q$ at Cartesian coordinate  $(x=0,y=0,z=0)$. 
	The image of this site inside the metal has charge $q$, and is located at $(0,0,-2r_i)$.  The Coulomb energy of interaction between the condensed hydronium ion located at $(2r_i,0,0)$ and all the functional groups, as well as their uniform neutralizing background is: 
	\begin{equation}\label{fund1}
	\begin{split}
	\beta \mu_s = -\frac{\lambda_B}{l}\sum_{n,m}\frac{1}{\sqrt{(n+\frac{m}{2}- \frac{2 r_{i}}{l})^2 + (\frac{\sqrt{3}}{2}m)^2} } +\\
	\frac{\lambda_B}{\gamma}\int\int_{A} dx\ dy \frac{1}{\sqrt{(x-2r_{i})^2+y^2}}\,.
	\end{split}
	\end{equation}
	On the other hand, the Coulomb interaction energy of the hydronium ion with its own image and the images of all the acid groups, as well as their neutralizing background is:
	\begin{equation}\label{fund2}
	\begin{split}
	\beta \mu_i = \frac{\lambda_B}{l} \sum_{n,m}\frac{1}{\sqrt{(n+\frac{m}{2}- \frac{2 r_{i}}{l})^2 + (\frac{\sqrt{3}}{2}m)^2 + (\frac{2  r_{i}}{l})^2}  } \\
	-\frac{\lambda_B}{\gamma} \int\int_{A} dx\ dy \frac{1}{\sqrt{(x- 2r_{i})^2+y^2+4r_{i}^2}}-\frac{\lambda_B}{4 r_{i}} \ .
	\end{split}
	\end{equation}
	The integral terms in Eqs.~\ref{fund1} and~\ref{fund2} are due to the interaction of hydronium with the respective neutralizing backgrounds of sites and image sites.  The last term in Eq.~\ref{fund2}  is due to the self interaction of the hydronium with its own image. The Eqs.~\ref{fund1} and \ref{fund2} converge very slowly. 
	In Appendix~\ref{Ap1} and \ref{Ap2} we present two efficient methods to perform these calculations.

	\section{Monte Carlo simulations}\label{A3}
	
\begin{figure}[t]
		\includegraphics[scale=0.3]{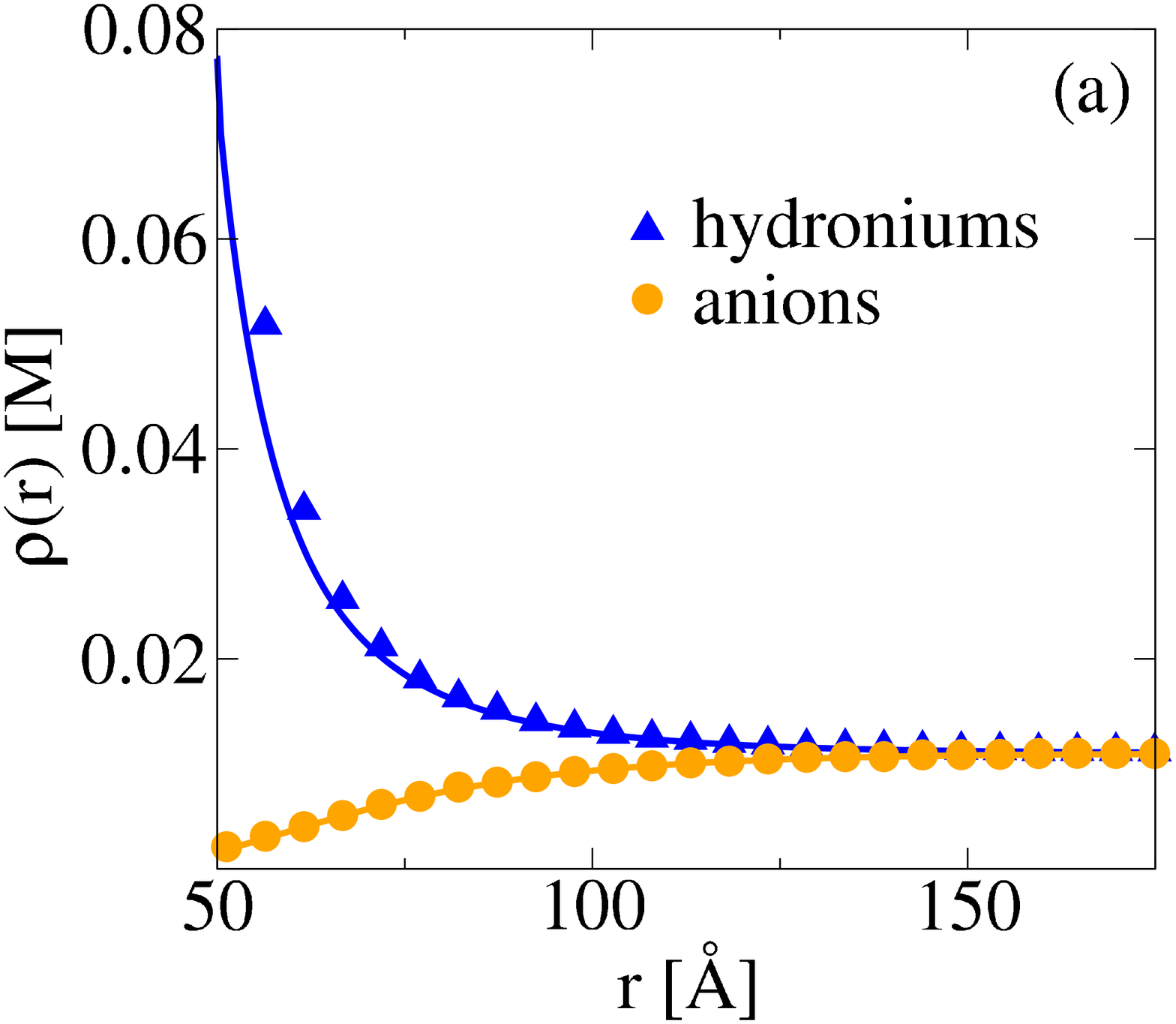}
		\includegraphics[scale=0.3]{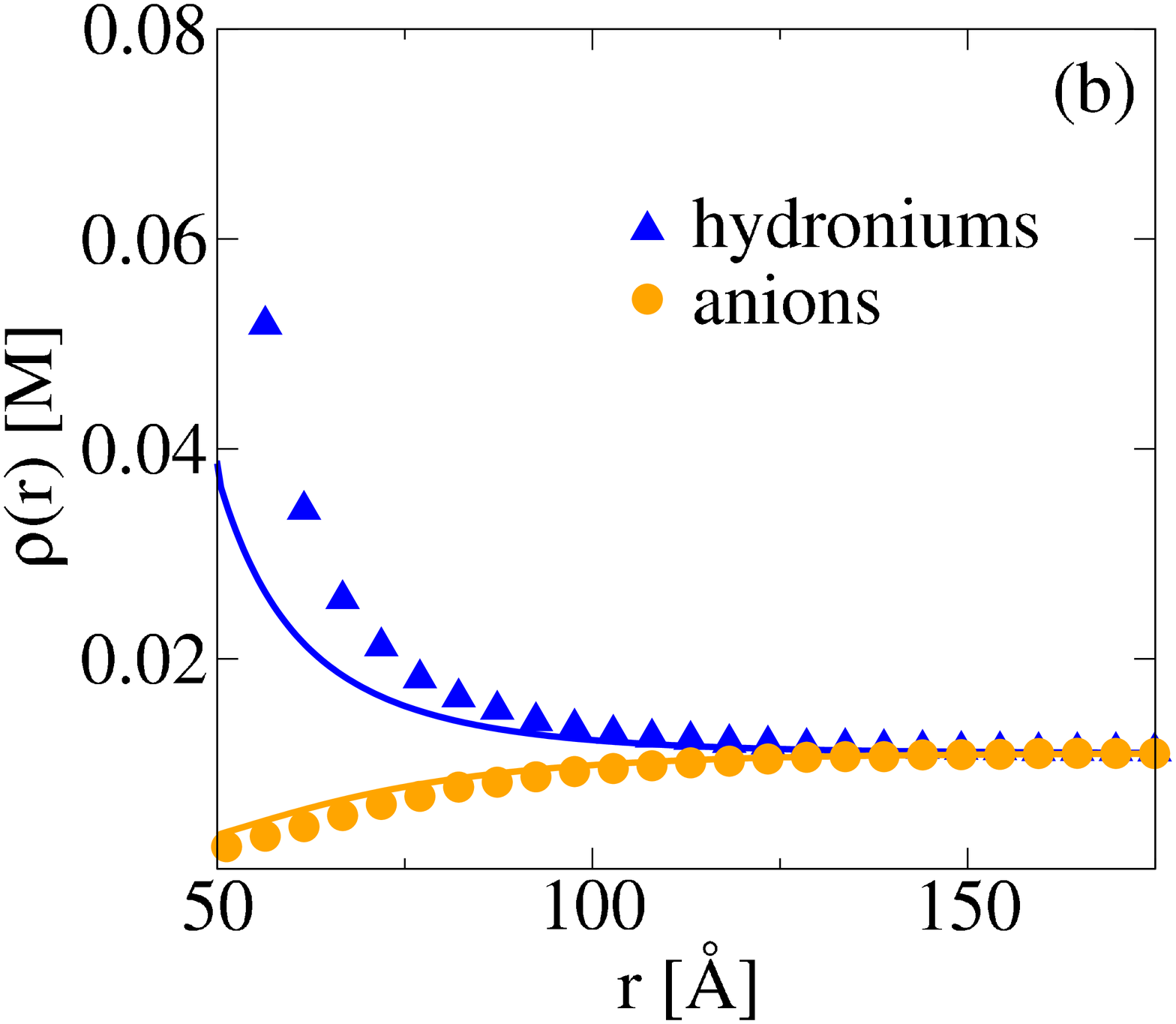}
		\caption{Ionic density profiles around a spherical metal nanoparticle with $N_s=100$ surface acid groups and radius $a=48$\AA. The radii of ions and of surface adsorption sites are $r_i=2$\AA. Comparison between MC simulations (symbols) and the CR theory (lines).  The reservoir has ${\rm pH}=1.95$ and no salt. (a) Using $K_{s}$ in PB. (b) Using $K_{bulk}$ in PB.}
		\label{figprof1}
		\includegraphics[scale=0.3]{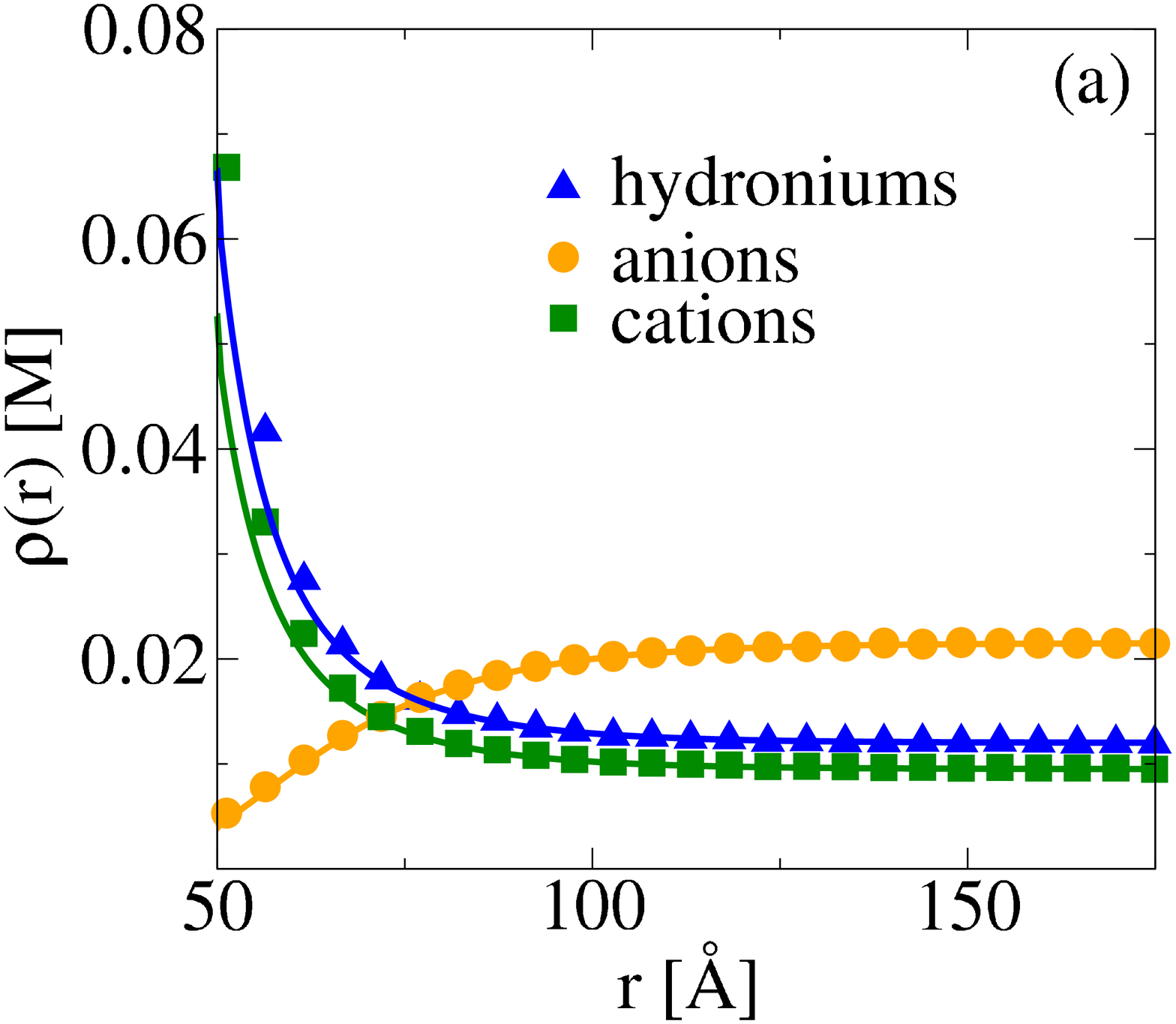}
		\includegraphics[scale=0.3]{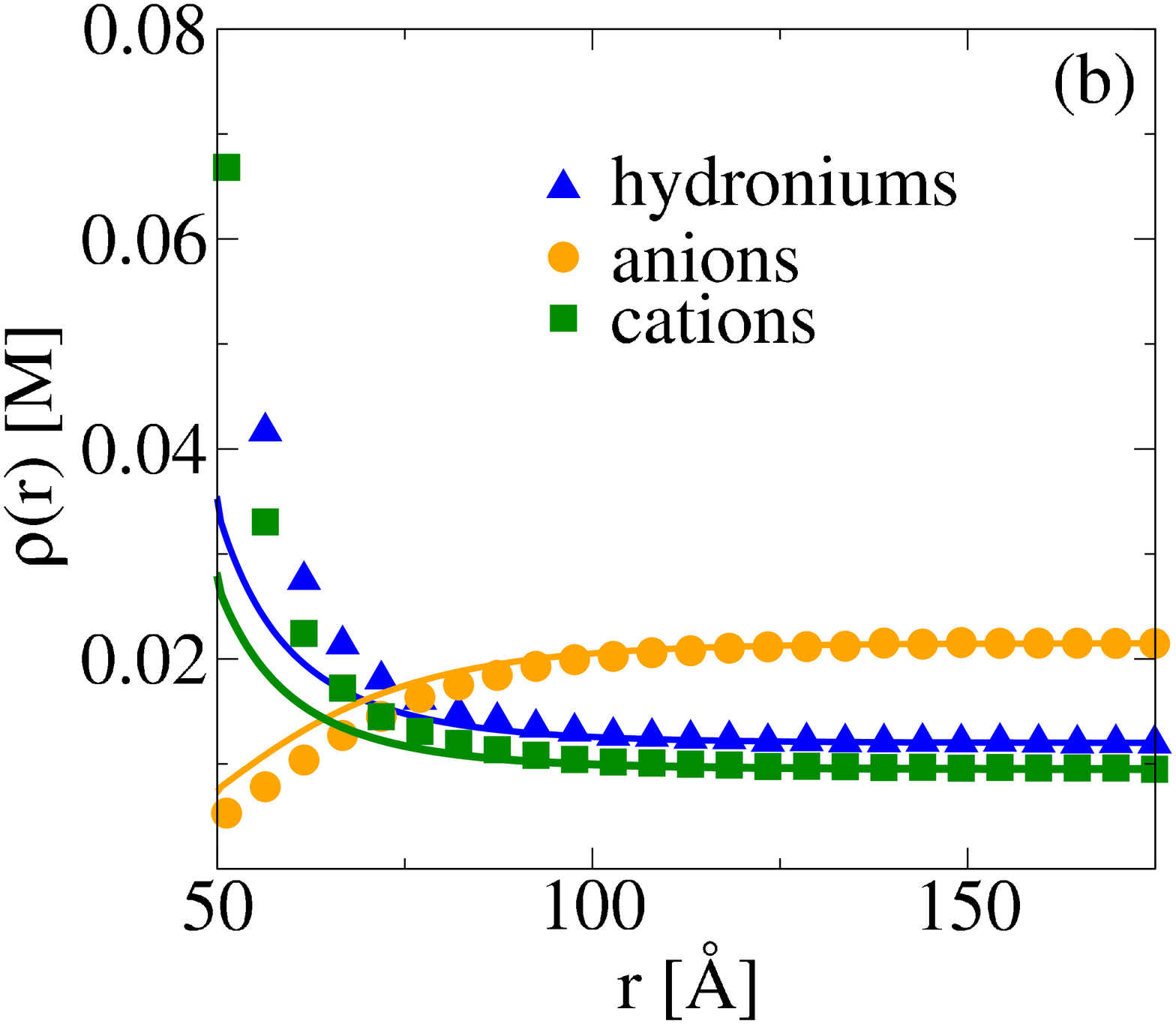}
		\caption{Ionic density profiles around spherical metal nanoparticle with  $N_s=100$ surface acid groups and radius $a=48$\AA. The radii of ions and adsorption sites are $r_i=2$\AA.  Comparison between MC simulations (symbols) and CR theory (lines). The reservoir has ${\rm pH}=1.92$ and salt concentration $9.5$mM. (a) Using $K_{s}$ in PB. (b) Using $K_{bulk}$ in PB.}
		\label{figprof2}
	\end{figure}

	In order to test the accuracy of the present theory we compare the ionic density profiles predicted by Eq. (\ref{eq1a}) with the results of Monte Carlo (MC) simulations. A spherical metal nanoparticle of radius $a$ and $N_s$ sticky adsorption sites of charge $-q$ and radius $r_i$, randomly distributed on its surface, is placed at the center of a spherical Wigner-Seitz cell of radius $R$. We consider a primitive model of electrolyte in which all ions have radius $r_i$, same as of the adsorption sites, and charge $\pm q$.  The system is in contact with a reservoir of strong acid and salt at respective concentrations,  $\rho_a$ and $\rho_s$.  The metal nature of nanoparticle is taken into account using the Green function, which accounts for the image charge~\cite{Joachim2018}.   The image charge of each ion is located at the corresponding inversion point inside the nanoparticle core. The association between the hydronium ions and the functional groups is taken into account using the Baxter sticky potential~\cite{bakhshandeh2019,bk2020},~$u_{Ba}$.  The electrostatic potential at position  ${\bf r}$ produced by a source located at position ${\bf r'}$ can be evaluated using the Green function of a conducting sphere~\cite{jack,dos2011,bkh2011,Joachim2018,BKH2018}:
	\begin{equation}\label{k1}
	G({\bf{r}},{\bf{r'}})=\frac{1}{4 \pi \epsilon_w \left|{\bf{r}}-{\bf{r'}}\right|}-\frac{a}{4 \pi \epsilon_w r' \left|{\bf{r}}-\frac{a^2}{r'^2}{\bf{r'}}\right|}+\frac{a}{4 \pi \epsilon_w  r' \left|{\bf{r}}\right|} .
	\end{equation}
	The first term is the result of the direct Coulomb interaction between an ion at position ${\bf r}$ and another ion at position ${\bf r'}$.  The second term is due to the interaction of the ion at position ${\bf r}$ with the image of the ion at ${\bf r'}$, which is placed at the inversion point inside the metal core  and has charge $-q a/r'$.  The last term is due to the interaction of the ion at ${\bf r}$ with the countercharge $+q a/r'$, placed at the center of the metal sphere to keep  the overall charge neutrality.  One can show that this construction leads to vanishing electric field inside the metal core and makes the particle equipotential~\cite{jack}. The total energy of the system can now be written as 
	\begin{equation}\label{k3}
	U=\sum_{i=1}^{N-1} \sum_{j=i+1}^{N}{q_i q_j G({\bf{r_i}},{\bf{r_j}})} +\frac{1}{2}\sum_j \bar G({\bf{r_j}},{\bf{r_j}}) +\sum_i u_{Ba}(\bold{r}_i),
	\end{equation}
	Where the $\bar G({\bf{r_j}},{\bf{r_j}})$ is the self interaction of an ion with its image inside the metal core and with the countercharge.  It is calculated using Eq. (\ref{k1}) without the direct Coulomb term.  The first sum in Eq.(\ref{k3}) runs over all the charged species inside the system, including the surface sites.  The second sum runs over all the ions inside the WS cell, and the last sum is for the sticky interaction 
	between surface sites and hydronium ions.
	The simulations are performed using the grand canonical Monte Carlo algorithm with $3\times 10^6$ MC steps for equilibration and $10^5$ steps for production.  The ionic density profiles obtained from the theory and from the MC simulations are presented in Figs.~\ref{figprof1}a and~\ref{figprof2}a.  As can be seen there is an excellent  agreement between the simulations and the theory.  For comparison we have also presented the density profiles obtained using the original NP approach in which the bulk association constant $K_{bulk}$ is used directly in the Eq. (\ref{Eq2}), instead of the renormalized surface equilibrium constant $K_s$, given by Eq. (\ref{ksur}).  As can be seen in Figs.~\ref{figprof1}b and~\ref{figprof2}b, the deviations between the original NP theory and the simulations are quite significant.  

	
	
	
	\section{Force between two metal nanoparticles}\label{A2}
	Consider the system represented in Fig.~\ref{fig1}:  two metal nanoparticles with $N_{s1}$ and $N_{s2}$  acid functional groups and effective radius $a_e$,  separated by the contact surface-to-contact surface distance $L$, inside an electrolyte-acid solution.  The concentration of  strong acid in the reservoir is $\rho_a=10^{-\rm{pH}}$, and of 1:1 salt is $\rho_s$.  To calculate the interaction force between the two nanoparticles we numerically solve the Poisson-Boltzmann equation in cylindrical coordinates $(\varrho,z)$, using the over-relaxation method~\cite{numericalrecipes}.  Since the metal core of each nanoparticle is an equipotential volume,  we apply the following boundary conditions: $\phi(\infty,z)=\phi(\varrho,\pm \infty)=0$, $\partial\phi(\varrho,z)/\partial \varrho |_{\varrho=0}=0$, $\phi|_{S_1}=\phi_1$ and $\phi|_{S_2}=\phi_2$, where $\phi_1$ and $\phi_2$ are the electrostatic potentials inside the nanoparticles 1 and 2, respectively. Our algorithm then performs a search for the potentials $\phi_1$ and $\phi_2$, such that the effective charge of each nanoparticle satisfies the CR boundary condition:
	\begin{equation}
	Q_{eff}^{(i)}=\frac{\epsilon_w}{4\pi} \int_S \mathrm{\bf E}\cdot d\mathrm{\bf S}=\frac{N_{s}^{(i)}}{(1+K_{s}\rho_a e^{-\beta q \phi_i})}
	\label{cr}
	\end{equation}
	where $i=1,2$; and the electric field, $\mathrm{\bf E}=-\nabla \phi(\varrho,z)$, is calculated on the contact surface $\mathrm{\bf S}$ of each nanoparticle.
	The electroosmotic stress tensor is given by
	\begin{equation}
	\Pi_{ij}=-p(\varrho,z)\delta_{ij}+\frac{\epsilon_w}{4\pi}\left[ E_i(\varrho, z)E_j(\varrho, z)-\frac{1}{2}E^2(\varrho, z)\delta_{ij}\right]
	\end{equation}
	where the kinetic pressure is $\beta p(\varrho, z)=(\rho_a+\rho_s)[e^{-\beta q \phi(\varrho, z)}+e^{\beta q \phi(\varrho, z)}]=2(\rho_a+\rho_s) \cosh[\beta q\phi(\varrho, z)]$.
	The electroosmotic force in $z$ direction felt by the nanoparticles is obtained by:
	\begin{equation}
	F=\int\hat{\mathbf z}\cdot {\mathbf\Pi}\cdot \hat{\mathbf n} dA,
	\end{equation}
	where the integration is performed over the surface of a cylinder enclosing one of the particles, see Fig. \ref{fig1},
	\begin{eqnarray}
	\MoveEqLeft[40] \beta F=2 \pi \int_0^{a_e} d\varrho \varrho \left\{2 (\rho_a+\rho_s) \left[ \cosh[\beta q\phi(\varrho,L/2)] - \cosh[\beta q\phi(\varrho,L/2+2a_e)\right]]\right . +\nonumber\\ 
	\MoveEqLeft[40] \frac{\beta\epsilon_w}{8 \pi} [ E_\varrho^2(\varrho, L/2) - E_z^2(\varrho, L/2) + \nonumber \\
	\MoveEqLeft[40]E_z^2(\varrho, L/2+2a_e) - E_\varrho^2(\varrho, L/2+2a_e)]\} + \nonumber \\
	\MoveEqLeft[40] 2\pi a_e \int_{L/2}^{L/2+2a_e} dz\ \frac{\beta\epsilon_w}{4\pi} E_\varrho(a_e, z)E_z(a_e, z)\ ,
	\label{eq2}
	\end{eqnarray}
	where the positive sign of the force signifies a repulsion. 
	\begin{figure}[t]
		\begin{center}
			\includegraphics[scale=0.3]{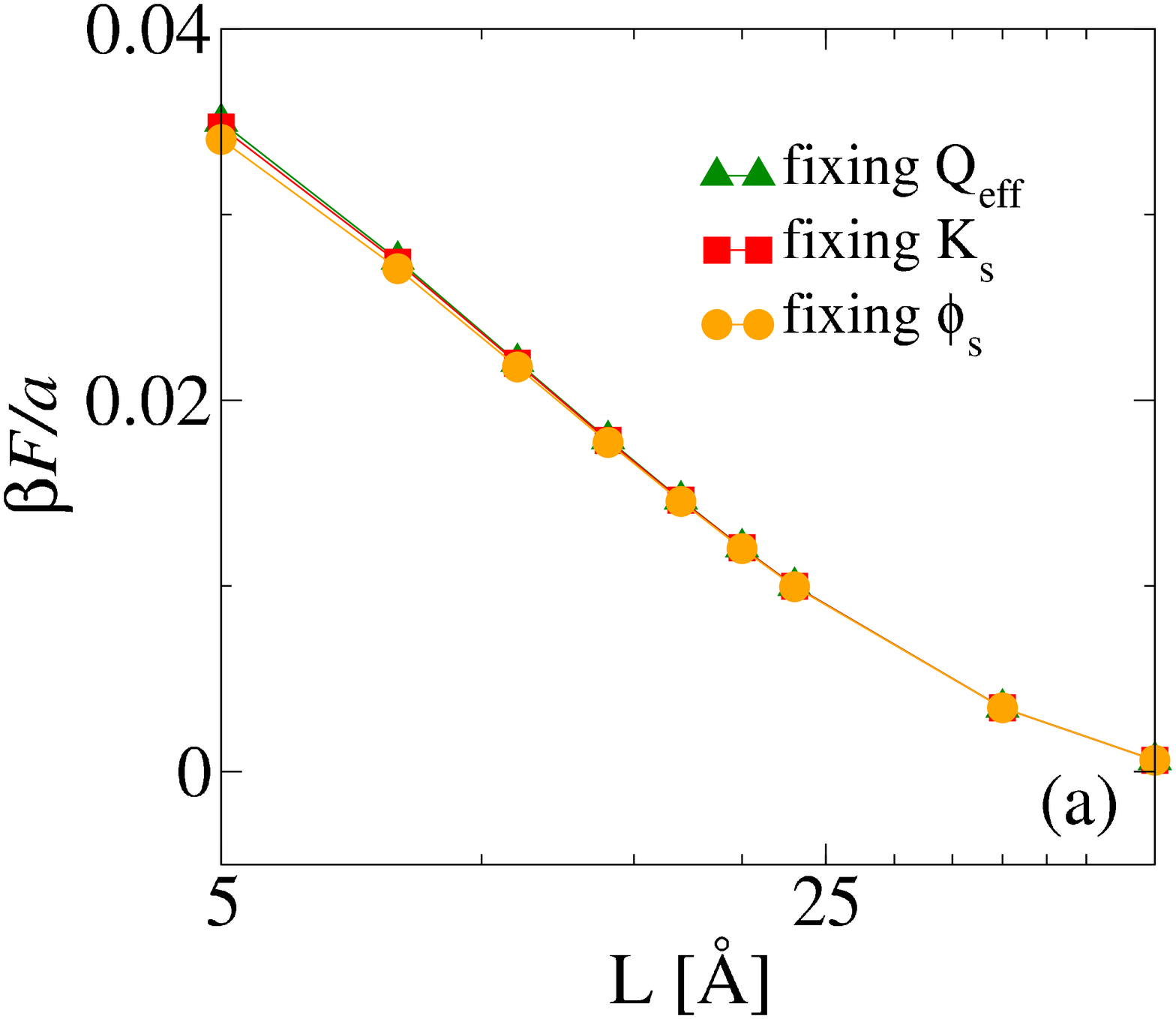}
			\includegraphics[scale=0.3]{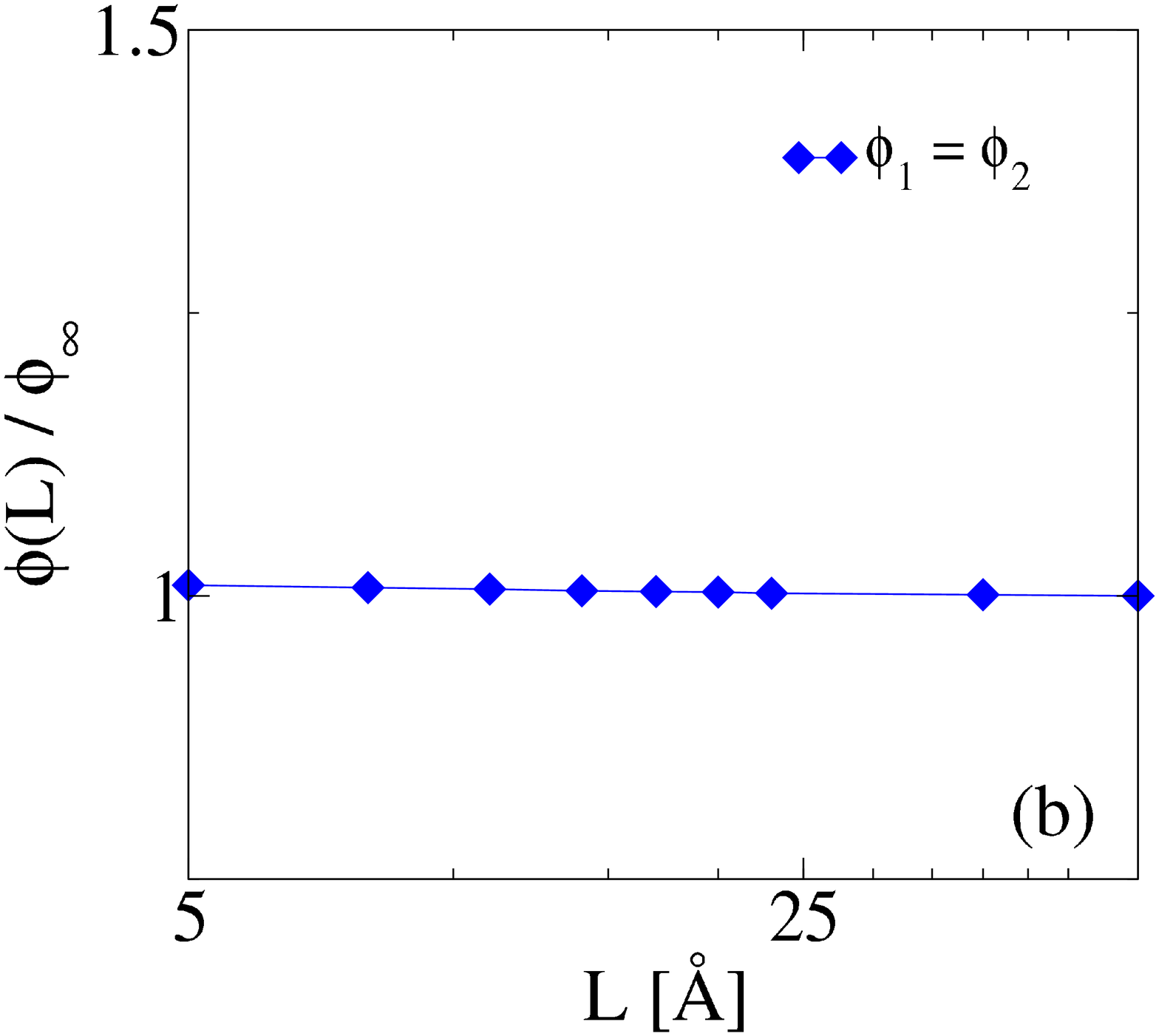}
		\end{center}
		\caption{(a) Force between two nanoparticles of radius $a=48$\AA\ as a function of surface to surface separation $L$ inside an acid solution at ${\rm pH}=5$ and salt concentration of  $20 $mM. The number of negatively charged sites on the two nanoparticle are $N_s^{(1)}=N_s^{(2)}=200$, respectively. The asymptotic  electrostatic potentials of the two particles are $\phi^{1}_\infty=\phi^{2}_\infty=-4.45 k_BT/q$ and the asymptotic charges are $Q^1_{eff}=Q^2_{eff}=-165q$. We present force calculated using 3 different boundary conditions: charge regulation boundary condition; fixed electrostatic potential boundary condition, in which the each particle electrostatic potential is kept the same as when the two particles are at infinite separation; and a fixed charge boundary condition, when the total charge on each particle is kept the same as when the particles are at infinite separation.
			(b) Electrostatic potentials $\phi_1(L)$ and $\phi_2(L)$ of nanoparticles with charge regulation, as a function of surface-to-surface separation.}
		\label{fig2}
	\end{figure}
	\begin{figure}[H]
		\begin{center}
			\includegraphics[scale=0.3]{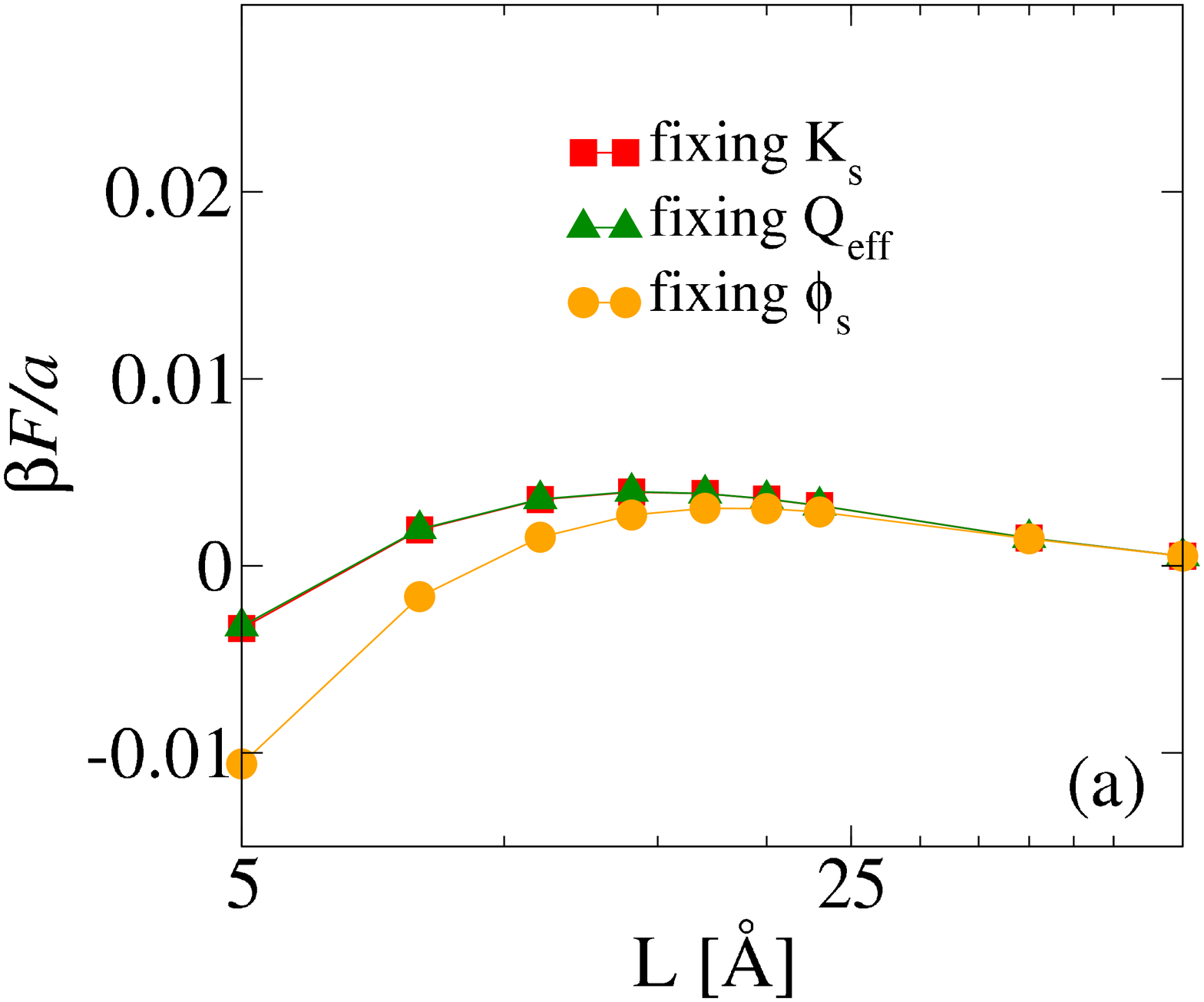}
			\includegraphics[scale=0.3]{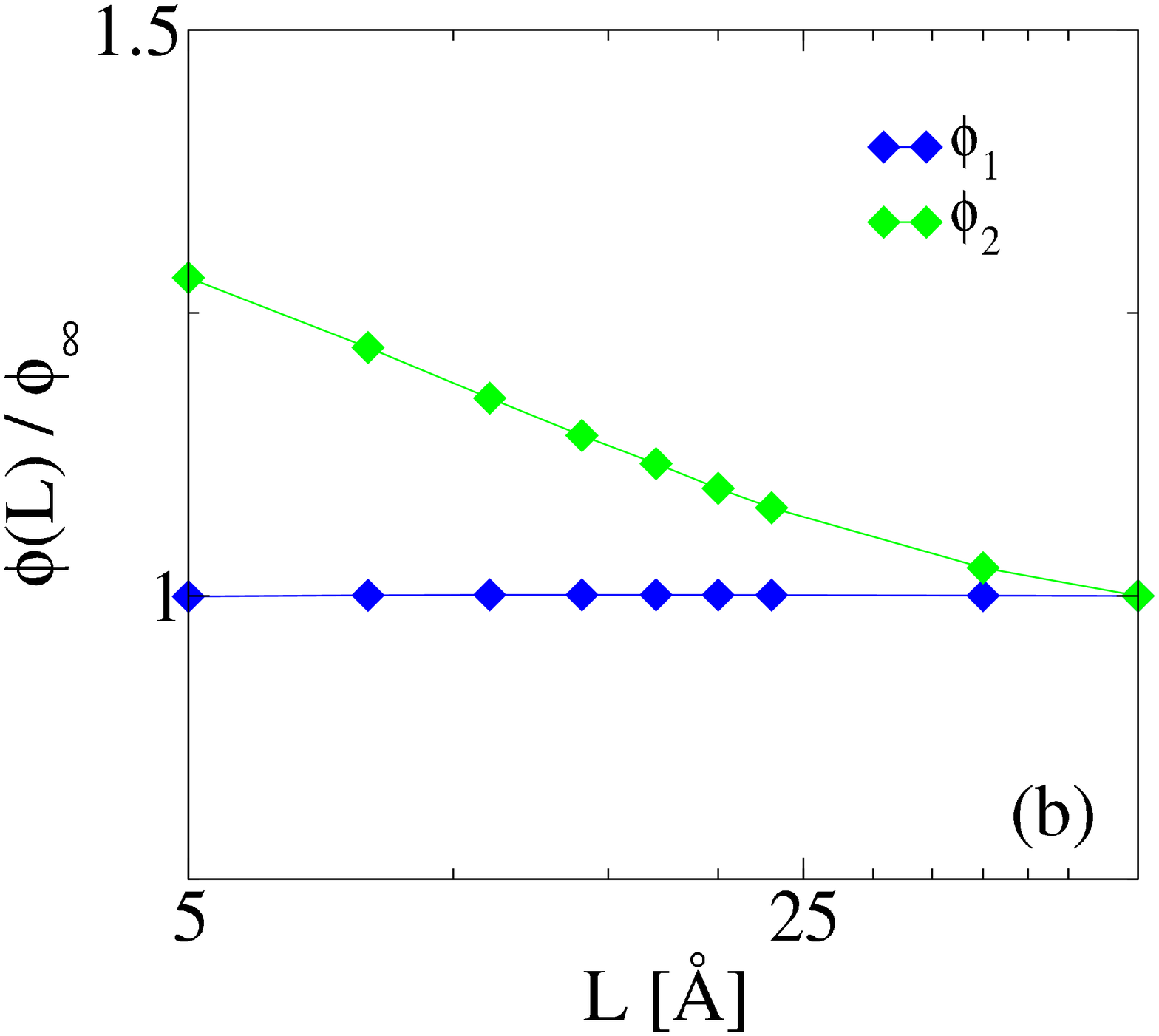}
		\end{center}
		\caption{(a) Force between two nanoparticles of radius $a=48$\AA\ as a function of surface to surface separation $L$. The ${\rm pH}=5$ while the salt concentration is $20$mM. The first particle has  $N_{s}^{(1)}=200$ negative surface functional groups and the second one has $N_{s}^{(2)}=30$. The asymptotic -- infinite separation between the particles --- values of the surface potential and effective surface charge are: $\phi^1_{\infty}=-4.45~k_BT/q$ and $\phi^2_{\infty}=-1.25 k_BT/q$ and for $Q^{(1)}_{eff}=-165q$ and $Q^{(2)}_{eff}=-29q$. We present force calculated using 3 different boundary conditions: charge regulation boundary condition; fixed electrostatic potential boundary condition, in which the each particle electrostatic potential is kept the same as when the two particles are at infinite separation; and a fixed charge boundary condition, when the total charge on each particle is kept the same as when the particles are at infinite separation. Note a very good agreement between CR boundary condition and the constant charge boundary condition, while the constant potential boundary condition significantly overestimates the like-charge attraction between the two nanoparticles at short separations.
			(b) Electrostatic surface potentials $\phi_1(L)$ and $\phi_2(L)$ of nanoparticles for CR boundary conditions, as a function of surface-to-surface separation. Note a strong variation of the electrostatic potentials at short distances, invalidating the often used constant potential boundary condition for calculating the interaction potential between the charge asymmetric nanoparticles.}
		\label{fig3}
	\end{figure} 
	
	\section{Results}\label{A4}
	
	We first study interaction between metal nanoparticles with the same number of surface functional groups inside an acid/salt solution, see Fig.~\ref{fig2}.    We explore three different boundary conditions:  CR boundary condition, Eq. (\ref{cr}); fixed electrostatic potential boundary condition --- in which each particle's electrostatic potential is kept the same as when the two particles are at infinite separation; and a fixed charge boundary condition, when the total charge on each particle is kept the same as when the particles are at infinite separation. For identical particles, we observe no significant effect of the boundary condition on the interaction force, Fig.~\ref{fig2}(a).  The insensitivity happens due to a very small variation of the  surface electrostatic potential with the particle separation, Fig.~\ref{fig2}(b).   This justifies the constant electrostatic potential boundary condition often used in the colloidal literature.  
	
	In Fig.~\ref{fig3}(a) we show the interaction force between particles with different number of surface adsorption sites. The other parameters are the same  as in the previous case. We note the appearance of like-charge attraction between the two metal nanoparticles at  small separations.   Such like-charge attraction was previously predicted for charge asymmetric metal nanoparticles based on the constant surface charge boundary condition~\cite{DosSantos2019}.  Indeed, we note almost a perfect agreement between CR boundary condition and a constant charge boundary condition, showing that the total charge on the metal nanoparticle surface does not vary  significantly with the separation between the nanoparticles.   On the other hand, the distribution of this charge over the surface of the metal core changes as the particles approach one another, affecting the electrostatic surface potential, see Fig.~\ref{fig3}(b).  While the surface charge is uniformly distributed on the two particles when they are far apart,  at close separation the charge on the weaker charged particle, redistributes in such as to induce a positive charge in the part of the metal core facing the other particle.  This leads to a like-charge attraction between the two metal nanoparticles, even though both are overall negatively charged.  We also see that for asymmetrically charged particles the constant electrostatic potential boundary condition overestimates the
	attraction between the two metal nanoparticles at short separations, see Fig.~\ref{fig3}(b), while the CR and constant charge boundary conditions remain in agreement.

	\section{Conclusions}\label{A5}
	
	We have presented a theoretical method for calculating the force between two charge regulating metal nanoparticles inside an electrolyte-acid solution.  Comparison between theory and simulations shows the importance of using the correct surface equilibrium constant when studying charge regulation of metal nanoparticles.   The bulk equilibrium constant must be renormalized to properly account for the discrete charge and steric effects at the particle surface.  
	
	Depending on the asymmetry in the number of acid functional groups on the two particles it is possible to obtain either repulsion or attraction between the two like-charged nanoparticles.   Our approach also shows that for metal nanoparticles it is always possible to replace the charge regulation boundary condition by a constant charge boundary condition.  On the other hand, the often used constant potential boundary works well for symmetric particles, but deviates significantly from the CR boundary condition for charge asymmetric particles, in particular in low salt electrolyte solutions.    
	
	Our calculations are based on the mean-field Poisson-Boltzmann equation, therefore we are not able to observe the Kirkwood-Shumaker attraction~\cite{kirkwood1952}, which appears close to the isoelectric point of the nanoparticles.  This attraction  results from the correlations between the associated hydronium ions on the two particle surfaces.  To explore this effect requires going beyond the mean-field theory.  This will be the subject of the future work.    Finally, the simulation results presented in the paper were obtained only for the individual particles inside a spherical WS cell.  To calculate the force between two charge regulated metal nanoparticles using MC simulations is significantly more difficult,  since it requires taking into account an infinite number of image charges. On the other hand, it is quite straightforward to obtain the interaction potential for non-polar particles using MC simulations~\cite{Thiago2012}, while the solution of PB equation for such particles is much more involved.  In the future work we will explore the difference in the interaction potential between polar and non-polar nanoparticle.  We will also attempt to develop numerical methods which will allow us to calculate the force between metal nanoparticles using MC simulations.   
	
	\section{Acknowledgments}
	
	This work was partially supported by the CNPq and FAPERGS.

	\section{Appendix}
	\subsection{Derivation of $\mu_s$}\label{Ap1}
	
\begin{figure}[H]
		\begin{center}
			\includegraphics[scale=0.65]{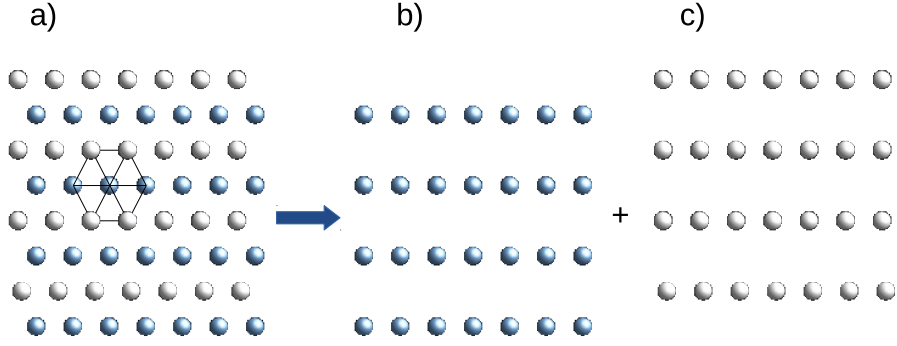} 
		\end{center}
		\caption{Triangular lattice as the sum of two square lattices.}
		\label{lattice}
	\end{figure}

	We neglect the curvature of the particle surface and consider a planar triangular lattice of adsorption sites of radius $r_i$ located at $z=0$.  The image 
	of the lattice is located at $z=-2r_i$, inside the metal core.	
	The triangular lattice can be decomposed into two simple rectangular sublattices, shifted with respect to one another by $1/2$ lattice spacing, in both $x$ and $y$ directions~\cite{vsamaj2012ground,vsamaj2012critical}, as is shown in Fig.~\ref{lattice}.
	We split $\mu_s$ into $\mu_{s_1}$ and $\mu_{s_2}$, where the indices 1 and 2 refer to the sublattices 1 and 2, respectively.
	The lattice vectors for the sublattice are: 
	\begin{equation}
	\begin{split}
	\boldsymbol{\mathrm{a}}_1 = l~{\boldsymbol{\hat{\mathrm x}}} \ \ \ \text{and} \ \ \ \boldsymbol{\mathrm{a}}_2 =  \Delta_1 l~{\boldsymbol{\hat{\mathrm y}}} \ ,
	\end{split}
	\label{vector1}
	\end{equation}
	where $\Delta_1=\sqrt{3}$. There  $\Delta_1$ and $\sigma$ are related as follows:
	\begin{equation}\label{sigma}
	\sigma= \frac{1}{ \abs{\boldsymbol{\mathrm{a}}_1 \times \boldsymbol{\mathrm{a}}_2}}= \frac{1}{l^2 \Delta_1}.
	\end{equation}
	The position of the hydronium which is in contact with the central site of the first sublattice located at (0,0,0), is assumed to be $(2r_{i},0,0)$.  The electrostatic interaction of hydronium with the adsorption sites of the first sublattice  and with their neutralizing background is:
	\begin{eqnarray}\label{de1}
	\beta \mu_{s_1} = -\frac{\lambda_B}{l}\sum_{n,m}{\frac{1}{\sqrt{(n -  \frac{2r_{i}}{l})^2 + \Delta_1^2~m^2}}}+
	\lambda_B \sigma \int_0^{R} dS \frac{1}{r} \ ,
	\end{eqnarray}
	where $R$ is the cutoff  distance imposed on both the sum and the integral.
	Using the Gamma function representation we can write~\cite{arfken1999} 
	\begin{equation} \label{gamma}
	\frac{1}{\sqrt{z}} = \frac{1}{\sqrt{\pi}} \int_0^{\infty} \frac{dt}{\sqrt{t}} e^{-zt} \ \ \ \text{for} \ \ \ z>0 \ ,
	\end{equation}
	and Eq. (\ref{de1}) can be written as~\cite{vsamaj2012ground,vsamaj2012critical,Leeuw}
	\begin{eqnarray}
	\beta \mu_{s_1} = -\frac{\lambda_B}{l}\sum_{n,m}{\frac{1}{\sqrt{\pi}} \int_0^{\infty} \frac{dt}{\sqrt{t}} e^{-[ (n -  2r_{i}/l)^2 + \Delta_1^2~m^2] t}}+
	\lambda_B \sigma \int_0^{R} dS \frac{1}{r} \ .
	\end{eqnarray}
	
	The interaction with the background can be written as:
	\begin{eqnarray}\label{dev6}
	\lambda_B \sigma \int_0^{R} dS \frac{1}{r}=\frac{2\pi\lambda_B}{l^2 \Delta_1}  \int_0^{R}dr\ r \frac{1}{r}= \nonumber\\
	\frac{2\sqrt{\pi}\lambda_B}{l \Delta_1}  \int_0^{\infty} \frac{dt}{\sqrt{t}} \int_0^{R}dr\ r e^{-t r^2}= \nonumber \\
	\frac{2\sqrt{\pi}\lambda_B}{l\Delta_1}  \int_0^{\infty} \frac{dt}{\sqrt{t}} \frac{(1-e^{-t R^2})}{2t} \ .
	\end{eqnarray}
	Taking the limit $R\rightarrow \infty$ in the above expression yields
	\begin{eqnarray}
	\frac{\sqrt{\pi}\lambda_B}{\Delta_1 l}  \int_0^{\infty} \frac{dt}{t\sqrt{t}} \ ,
	\end{eqnarray}
	and the expression for $\mu_{s_1}$ becomes
	\begin{eqnarray}
	\beta \mu_{s_1} = -\frac{\lambda_B}{l}\frac{1}{\sqrt{\pi}} \int_0^{\infty} \frac{dt}{\sqrt{t}}\left[ \sum_{n,m} e^{-[ (n -  2r_{i}/l)^2 + \Delta_1^2~m^2] t}-
	\frac{\pi}{\Delta_1 t}\right] \ .
	\end{eqnarray}
	Splitting the integral into intervals $t=0\text{ to }t=\pi$ and $t=\pi\text{ to }t=\infty$, and changing variables $\pi^2/t \rightarrow u$ in the second interval, we obtain
	\begin{eqnarray}
	\label{a9}
	\beta \mu_{s_1} = -\frac{\lambda_B}{l}\frac{1}{\sqrt{\pi}} \int_0^{\pi} \frac{dt}{\sqrt{t}}\left[ \sum_{n,m} e^{-[ (n -  2r_{i}/l)^2 + \Delta_1^2~m^2] t}-
	\frac{\pi}{\Delta_1 t}\right] \\ \nonumber
	-\frac{\lambda_B\sqrt{\pi}}{l} \int_{0}^{\pi} \frac{du}{u^{3/2}}\left[ \sum_{n,m} e^{-[ (n -  2r_{i}/l)^2 + \Delta_1^2~m^2] \pi^2/u}-
	\frac{u}{\Delta_1 \pi}\right]  \ .
	\end{eqnarray}
	Using Poisson sum rule\cite{gasquet2013}
	\begin{equation}
	\sum_{n=-\infty}^{\infty} \mathrm{e}^{-(n+\omega)^2 \gamma t} = \sqrt{\frac{\pi}{\gamma t}}
	\sum_{n=-\infty}^{\infty} \mathrm{e}^{\frac{-\pi^2 n^2}{\gamma t}} \cos{(2\pi n\omega)}
	\label{eq:Poisson}
	\end{equation} 
	in first term of Eq. (\ref{a9}) we obtain
	\begin{eqnarray}
	\beta \mu_{s_1} = -\frac{\lambda_B \sqrt{\pi}}{l\Delta_1} \int_0^{\pi} \frac{dt}{t^{3/2}}\left[ \sum_{n,m} \cos{(4\pi n r_{i}/l)}\mathrm{e}^{-(n^2+m^2/\Delta_1^2)\pi^2/t} - 1 \right] \\ \nonumber
	-\frac{\lambda_B\sqrt{\pi}}{l} \int_{0}^{\pi} \frac{du}{u^{3/2}}\left[ \sum_{n,m} e^{-[ (n -  2r_{i}/l)^2 + \Delta_1^2~m^2] \pi^2/u}-
	\frac{u}{\Delta_1 \pi}\right] \ .
	\end{eqnarray}
	The $(n=0,m=0)$ term of first sum above cancels with the background term, resulting in
	\begin{eqnarray}
	\beta \mu_{s_1} = -\frac{\lambda_B \sqrt{\pi}}{l\Delta_1} \int_0^{\pi} \frac{dt}{t^{3/2}}\left[ \sum'_{n,m} \cos{(4\pi n r_{i}/l)}\mathrm{e}^{-(n^2+m^2/\Delta_1^2)\pi^2/t}\right] \\ \nonumber
	-\frac{\lambda_B\sqrt{\pi}}{l} \int_{0}^{\pi} \frac{du}{u^{3/2}}\left[ \sum_{n,m} e^{-[ (n -  2r_{i}/l)^2 + \Delta_1^2~m^2] \pi^2/u}-
	\frac{u}{\Delta_1 \pi}\right] \ ,
	\end{eqnarray}
	where the prime above the summation indicates the exclusion of $(n=0,m=0)$ term. Performing the integration we obtain,
	\begin{eqnarray}
	\beta \mu_{s_1} = -\frac{\lambda_B \sqrt{\pi}}{l\Delta_1} \left[ \sum'_{n,m} \frac{\text{erfc}{\left(\sqrt{\pi(n^2+m^2/\Delta_1^2)}\right)}}{\sqrt{\pi(n^2+m^2/\Delta_1^2)}}\cos{(4\pi n r_{i}/l)}\right] \\ \nonumber
	-\frac{\lambda_B\sqrt{\pi}}{l} \left[ \sum_{n,m} \frac{\text{erfc}{\left( \sqrt{\pi[ (n -  2r_{i}/l)^2 + \Delta_1^2~m^2]} \right)}}{\sqrt{\pi[ (n -  2r_{i}/l)^2 + \Delta_1^2~m^2]}}-\frac{2 }{\Delta_1 \sqrt{\pi} }\right] \ .
	\end{eqnarray}
	Both sums converge very fast.
	Similar procedure can be used to calculate $\mu_{s_2}$ after noting that second sublattice   is shifted with respect to the first sublattice by $l/2$ and $\Delta_1 l/2$ in $x$ and $y$ directions, respectively. We obtain:
	\begin{eqnarray}
	\beta \mu_{s_2} = -\frac{\lambda_B \sqrt{\pi}}{l\Delta_1} \left[ \sum'_{n,m} \frac{\text{erfc}{\left(\sqrt{\pi(n^2+m^2/\Delta_1^2)}\right)}}{\sqrt{\pi(n^2+m^2/\Delta_1^2)}}\cos{\left(\frac{4\pi n r_{i}}{l}- \pi n\right)}\cos{(\pi m)}\right] \\ \nonumber
	-\frac{\lambda_B\sqrt{\pi}}{l} \left[ \sum_{n,m} \frac{\text{erfc}{\left( \sqrt{\pi[ (n -  2r_{i}/l - 1/2)^2 + \Delta_1^2~(m-1/2)^2]} \right)}}{\sqrt{\pi[ (n -  2r_{i}/l - 1/2)^2 + \Delta_1^2~(m-1/2)^2]}}-\frac{2}{\Delta_1 \sqrt{\pi}}\right] \ .
	\end{eqnarray}
	\subsection{Derivation of $\mu_i$}\label{Ap2}
	
	To calculate the interaction energy between the ion and the images,  $\mu_i$, we can use a similar procedure to the one described above.  However, the expressions become somewhat more involved.  Therefore, here we provide an alternative approach which also leads to a rapidly converging expression for the value of  $\mu_i$.  Consider a triangular array of image sites located at $z=0$.  For convenience of calculations we have shifted the plane of image sites from $z=-2 r_i$ to $z=0$, so that the adsorbed hydronium ion is now at the coordinate $(2r_i,0, 2r_i)$. The electrostatic potential produced by these sites satisfies the Poisson equation
	\begin{equation}
	\nabla^2 G(\boldsymbol{r}) = -\frac{4 \pi q}{\epsilon_w}  \sum_{n,m} \delta(z)\delta(\boldsymbol{\alpha} -n \boldsymbol{\mathrm{a}}_1-m \boldsymbol{\mathrm{a}}_2) \ ,
	\label{Green}
	\end{equation}
	where $\boldsymbol{\alpha}=  \boldsymbol{{\mathrm x}}~ \boldsymbol{\hat{\mathrm x}}+\boldsymbol{{\mathrm y}}~ \boldsymbol{\hat{\mathrm y}}$, and $\boldsymbol{\mathrm{a}}_1$ and $\boldsymbol{\mathrm{a}}_2$ are given by Eq.~(\ref{vector}).  The source term of the 
	Poisson equation can be
	rewritten using the Fourier representation of the periodic delta function~\cite{bk2020}
	\begin{equation}
	\begin{split}
	\sum_{n,m} \delta(\boldsymbol{\alpha} -n \boldsymbol{\mathrm a}_1-m \boldsymbol{\mathrm a}_2)=
	\frac{1}{\gamma}  \sum_{n,m} \mathrm{e}^{i \boldsymbol{\alpha} \cdot (n\boldsymbol{\mathrm b}_1  + m\boldsymbol{\mathrm b}_2)} \ ,
	\end{split}
	\label{v3}
	\end{equation}
	where $\gamma = \frac{\sqrt{3}}{2} l^2$ is the unit cell area of the triangular lattice, while the reciprocal lattice vectors are:
	\begin{equation}
	\begin{split}
	\boldsymbol{\mathrm b}_1 = \frac{2 \pi}{l} \left({\boldsymbol{\hat{\mathrm x}}} - \frac{{\boldsymbol{\hat{\mathrm y}}}}{\sqrt{3}}\right) \ \ \ \text{and} \ \ \ 
	\boldsymbol{\mathrm b}_2 = \frac{2 \pi}{l} \left( \frac{2 \boldsymbol{\hat{\mathrm y}}}{\sqrt{3}}\right) \ .
	\end{split}
	\label{v2}
	\end{equation}
	The Green function can be written as~\cite{Malossi2020,dos17}
	\begin{equation}
	\begin{split}
	G(\boldsymbol{r}) = \frac{1}{\gamma} \sum_{n,m} g_{n,m}(z) \mathrm{e}^{i \boldsymbol{\alpha} \cdot (n \boldsymbol{\mathrm b}_1 + m \boldsymbol{\mathrm b}_2)} \ .
	\end{split}
	\label{v4}
	\end{equation}
	Using Eq.~\ref{v4} in Eq.~(\ref{Green}) we obtain
	\begin{equation}
	\begin{split}
	\frac{\partial^2 g_{n,m}(z)}{\partial z^2} -k^2 g_{n,m}(z)=-\frac{4 \pi q}{\epsilon_w}\delta(z) \ ,
	\end{split}
	\label{v6}
	\end{equation}
	where $k$ is given by
	\begin{equation}
	\begin{split}
	k = \sqrt{\frac{4 \pi^2}{l^2}\left(n^2 +\left(\frac{2 m }{\sqrt{3}} -\frac{n}{\sqrt{3}}\right)^2\right)} \ .
	\end{split}
	\end{equation}
	Integrating Eq. (\ref{v6}) once over $z$, and taking the limit $z \rightarrow 0$ from both sides,  we obtain
	\begin{equation}  
	g^{\prime}_{n,m}(0^+)-g^{\prime}_{n,m}(0^-) =-\frac{4 \pi q}{\epsilon_w} \ ,
	\label{v8}  
	\end{equation} 
	yielding
	\begin{equation}
	\begin{split}
	G(\boldsymbol{r}) = \frac{2 \pi q}{\gamma \epsilon_w} \sum_{n,m} \frac{\mathrm{e}^{-k\abs{z}}}{k}\cos{\left[\frac{2 \pi}{l}\left(n x+ \frac{1}{\sqrt{3}}(2 y m - y n) \right)\right]} \ .
	\end{split}
	\label{v10}
	\end{equation}
	The diverging $(n=0,m=0)$ term is canceled if a neutralizing background is introduced in the source term of the Poisson equation (\ref{Green}).  For the hydronium located located at $(2r_{i}, 0, 2r_{i})$, 
	the interaction energy with the images of the adsorption sites is $q G(2 r_{i}, 0, 2 r_{i})$.  The expression for $\mu_i$ can now be written in terms of a rapidly converging sum
	\begin{equation}\label{mu2}
	\begin{split}
	\beta \mu_i = \frac{2 \pi \lambda_B}{\gamma} \sum'_{n,m} \frac{\mathrm{e}^{-2kr_{i}}}{k}\cos{\left(\frac{4 \pi r_{i}}{l} n \right)} - \frac{\lambda_B}{4 r_{i}} \ ,
	\end{split}
	\end{equation}
	where prime over the sum signifies that the $(n=0, m=0)$ term is excluded from the summation.  The last term of Eq. (\ref{mu2}) due to the interaction of a hydronium ion with its own image.
	
\bibliographystyle{ieeetr}
\bibliography{ref}

\end{document}